\documentclass{Interspeech}

\interspeechcameraready
\title{Fast-VGAN: Lightweight Voice Conversion  with Explicit Control \\ of F0 and Duration Parameters}
   
\author[]{Mathilde}{Abrassart}
\author[]{Nicolas}{Obin}
\author[]{Axel}{Roebel}

\affiliation{STMS Lab}{IRCAM, CNRS, Sorbonne Université}{Paris, France}
\email{\{abrassart,obin,roebel\}@ircam.fr}

\usepackage{comment}
\usepackage{lipsum}
\usepackage{svg}
\usepackage[finalnew]{trackchanges}
\usepackage{adjustbox}
\usepackage{bm}
\usepackage{graphicx}
\usepackage{subcaption}
\usepackage{placeins}
\usepackage{float}

\addeditor{NO} 
\addeditor{AR} 
\addeditor{MA} 

\begin{document}

\maketitle

\begin{abstract}
    
    
    

    Precise control over speech characteristics, such as pitch, duration, and speech rate, remains a significant challenge in the field of voice conversion. The ability to manipulate parameters like pitch and syllable rate is an important element for effective identity conversion, but can also be used independently for voice transformation, achieving goals that were historically addressed by vocoder-based methods. 
    In this work, we explore a convolutional neural network-based approach that aims to provide means for modifying fundamental frequency (F0), phoneme sequences, intensity, and speaker identity. Rather than relying on disentanglement techniques, our model is explicitly conditioned on these factors to generate mel spectrograms, which are then converted into waveforms using a universal neural vocoder. Accordingly, during inference, F0 contours, phoneme sequences, and speaker embeddings can be freely adjusted, allowing for intuitively controlled voice transformations.
    We evaluate our approach on speaker conversion and expressive speech tasks using both perceptual and objective metrics. The results suggest that the proposed method offers substantial flexibility, while maintaining high intelligibility and speaker similarity. 

\end{abstract}

\section{Introduction}

Voice conversion (VC) aims to transform a source speaker’s voice to sound like that of a target speaker, while preserving the linguistic content \cite{bargum2024reimagining,sisman2020overview}. By definition, this task requires altering vocal identity, which raises fundamental questions about what features define a speaker’s voice and how they should be manipulated. 
Many systems treat the prosody characteristics as an implicit attribute of the speaker's identity~\cite{qian2019autovc,popov2022diffvc,guo2023quickvc}. They do not explicitly represent prosody-related features. This has the advantage of ensuring that the prosody will be coherent with the target speaker. \change{A particular}{An alternative} approach consists of preserving the source prosody. This strategy is notably interesting for voice reenactment in cinema production \cite{bous2022voice}.
Recently, the independent control of prosody for voice conversion models has received increasing interest \cite{byun2023highlycontrollablediffvc,kashkin2023hifi,chen2023controlvc}. These latter approaches enable explicit control over various prosodic dimensions (e.g., pitch, intensity, speaking rate), allowing users \change{to steer expressive style}{to control the prosody} independently of either source or target. In the present study, we are particularly interested in the detailed manipulation of prosody features in voice conversion applications. To this end, we introduce \textbf{Fast-VGAN}\footnote{A demo page of Fast-VGAN is available at the following link: \url{https://abrassartm.github.io/Fast-VGAN}}, a non-autoregressive, fully convolutional, GAN-based voice conversion model that maps high-level features (F0, intensity, phoneme embeddings) into mel-spectrograms while preserving speaker identity through speaker embeddings. Beyond voice identity conversion, the fine-grained control over F0 and duration in voice conversion is essential to locally and dynamically manipulate the expressivity, which is mostly conveyed by means of prosody (F0, intensity, duration, and voice quality \cite{campbell2003voice}. Disentangling prosody and speaker identity enables not only a more versatile and expressive control of voice conversion, but also facilitates the systematic study of how prosodic and timbral cues separately contribute to the perception of speaker identity \cite{benaroyavc2923}. Unlike diffusion-based systems, Fast-VGAN enables fast and lightweight inference, making it suitable for real-time or resource-constrained applications \cite{donahue2021endtoend}. Crucially, we aim to allow explicit control of prosodic dimensions (including speech rate) at inference time, without requiring expressive training data, thereby facilitating expressive voice conversion from neutral inputs. For instance, increasing pitch variability, often referred to as pitch range or ambitus, has been associated with more emotionally expressive speech, while modifying speaking rate through temporal expansion or compression can influence the perception of speaking style and intent \cite{gengembre2024disentangling, deng2024learning, qu2023disentangling, sivaprasad2021emotional}. Understanding how these factors influence VC outcomes is essential for applications such as personalized assistants, emotional speech generation, and dubbing for multimedia content \cite{rana2024advancements, lee2019robust}. 
It is well known that additional expressiveness comes at the cost of generalization or speaker fidelity \cite{tjandra2020unsupervised, wang2020learning}. Accordingly, we will investigate the relationship between such controls and the signal quality, including the perceptual impact on speaker similarity.  \\

\vspace{-0.15cm}
The remainder of this paper is organized as follows. Section~\ref{sec:related_works} provides an overview of related work in voice conversion, with a particular focus on prosodic modeling and controllable generation. Section~\ref{sec:proposed_method} details the architecture of the proposed Fast-VGAN system, including its input representations, conditioning mechanisms, and adversarial training framework. In Section~\ref{sec:experiments}, we describe the experimental protocol, including datasets, and the baseline models used for comparison. Section~\ref{sec:evaluations} presents both objective and subjective evaluations, designed to assess speaker similarity, intelligibility, and expressiveness across different conversion settings. Finally, Section~\ref{sec:conclusion} summarizes our findings. 

\vspace{-0.15cm}

\section{Related Works}
\label{sec:related_works}

\vspace{-0.1cm}

We review key areas relevant to our work, including recent developments in voice conversion models and prosodic control strategies, the role of feature representations in enabling fine-grained control, and the application of GAN-based approaches for high-quality and efficient speech synthesis.

\subsection{Voice Conversion Models and Prosodic Control}

Traditional VC methods primarily focused on modifying spectral features to achieve conversion. Recent approaches have shifted towards more comprehensive models that incorporate prosodic features for enhanced expressiveness and control.
HiFi-VC~\cite{kashkin2023hifi} proposes an encoder-decoder framework for any-to-any voice conversion, relying on bottleneck features from a pre-trained Conformer ASR model~\cite{gulati2020conformer} to extract linguistic content. To complement the limited prosodic capacity of ASR features, it incorporates a dedicated F0 encoder based on the WORLD vocoder~\cite{morise2016world}. A speaker embedding network~\cite{huang2021far} conditions the decoder, which unifies spectrogram generation and waveform synthesis via a conditional HiFi-GAN~\cite{kong2020hifi}. While HiFi-VC excels at audio quality and zero-shot conversion, it lacks fine-grained control over prosodic attributes such as pitch range or speech rate, limiting its expressiveness~\cite{du2021disentanglement, deng2024learning}.
ControlVC~\cite{chen2023controlvc} addresses this limitation by enabling time-varying controls on pitch and speed. It utilizes pre-trained encoders to generate pitch and linguistic embeddings, combined and converted to speech using a vocoder. Speed control is achieved through TD-PSOLA pre-processing~\cite{charpentier1986diphone}, while pitch control is managed by manipulating the pitch contour before feeding it into the encoder. Despite these advancements, ControlVC's reliance on pre-trained models like HuBERT~\cite{hsu2021hubert} for linguistic embeddings may limit the granularity of control over prosodic features. Moreover, the overall architecture remains relatively complex, combining several external modules and processing stages. 

\subsection{Feature Representations for Enhanced Control}
The choice of feature representations can play a crucial role in the controllability and expressiveness of VC systems. Unlike latent representations derived from large pre-trained models such as wav2vec~\cite{baevski2020wav2vec}, explicit and interpretable features like fundamental frequency (F0), phoneme durations, and intensity offer more direct control over prosodic attributes. Aligned phoneme sequences, in particular, provide a temporally precise and interpretable structure for manipulating speech rate. Numerous studies have explored different techniques for F0 extraction. Classical algorithms such as WORLD~\cite{morise2016world} and DIO~\cite{morise2009fast} have been widely used in VC and TTS pipelines. More recently, neural pitch estimators such as CREPE~\cite{kim2018crepe} and convolutional architectures~\cite{ardaillon2019fully} have demonstrated improved robustness and resolution.

\subsection{GAN-Based Approaches in Voice Conversion}
Generative Adversarial Networks (GANs) have been widely adopted in voice conversion (VC) for their ability to generate high-fidelity and natural-sounding speech. Pioneering works such as CycleGAN-VC \cite{kaneko2017parallel,ferrovc2020}, StarGAN-VC~\cite{kaneko2019starganvc} and its successor StarGAN-VC2~\cite{kaneko2020starganv2} enabled many-to-many VC without parallel data using a single GAN-based framework. Similarly, models like NVC-Net~\cite{qian2020nvcnet} and AVQVC~\cite{tang2022avqvc} demonstrated that adversarial training enhances voice naturalness and speaker similarity. More recently, HiFi-GAN~\cite{kong2020hifi} has become a standard neural vocoder used in conjunction with GAN-based decoders for both TTS and VC. Unlike diffusion-based models~\cite{popov2021grad, popov2022diffvc, choi2024dddm}, which require iterative sampling and incur high computational costs, GAN-based models enable real-time inference with low latency. Furthermore, adversarial objectives help overcome the limitations of traditional reconstruction losses, which often lead to over-smoothed or overly averaged spectrograms~\cite{kaneko2017parallel}.

\begin{figure*}[ht]
    \centering
\begin{subfigure}[b]{0.48\textwidth}
    \centering
\includegraphics[width=1\linewidth]{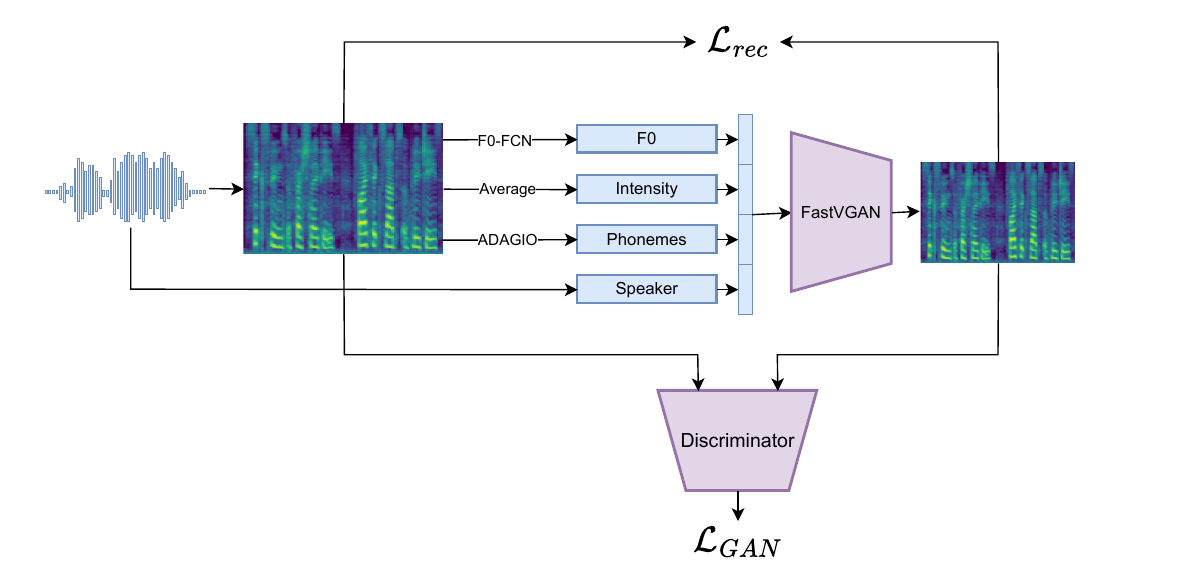}
    \caption{Fast-VGAN training phase}
\label{fig:fastvg-glob}
\end{subfigure}
\begin{subfigure}[b]{0.48\textwidth}
\centering
\includegraphics[width=1\linewidth]{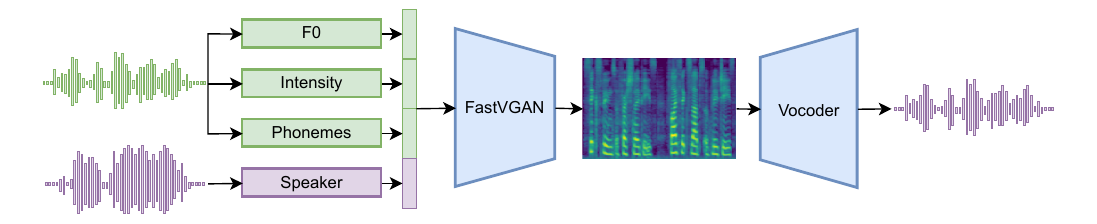}
\vspace{1.5cm}
    \caption{Fast-VGAN inference phase}
 \label{fig:fastvg-inf}
\end{subfigure}
\caption{Fast-VGAN architecture. During training: the $\mathcal{L}_{rec}$ loss is the RMSE loss and the $\mathcal{L}_{GAN}$ loss is the MSE loss. The weighting is 1 for $\mathcal{L}_{rec}$ and 0.5 for $\mathcal{L}_{GAN}$.}
\vspace{-0.25cm}
\end{figure*}

\section{Fast-VGAN Voice Conversion}
\label{sec:proposed_method}

In this section, we detail the rationale behind our architectural choices. Our goal is to design a model capable of transferring a specific utterance from one speaker to another speaker’s timbre, while relying on high-level control features. To facilitate training on limited data and ensure a lightweight implementation, we impose constraints on the model’s parameter count and structural complexity. We introduce Fast-VGAN, a lightweight architecture designed to resynthesize mel-spectrograms from interpretable conditioning inputs, including F0, intensity, phoneme sequence, and speaker identity. We describe the methodology used to extract and process these conditioning features.

\subsection{Overview}

Fast-VGAN employs a series of 2D convolutional layers tailored to process temporally aligned sequences of F0, intensity, and phoneme representations. The architecture is designed to map these high-level temporal features into a target mel-spectrogram. To enable timbre control in a multi-speaker setting, we incorporate an additional conditioning input: a speaker identity embedding. This enables the model to account for and reproduce distinct vocal timbres associated with different speakers in the training corpus. A high-level overview of the system architecture is presented in Figure \ref{fig:fastvg-glob}.
The spectrograms generated by Fast-VGAN are subsequently fed into the MBExWN vocoder \cite{roebel2022neural}, a universal neural vocoder used to synthesize the final audio waveform. A well-known challenge in spectrogram-based synthesis is that reconstruction losses such as RMSE tend to average out fine-grained details during training, leading to over-smoothed outputs that lack naturalness \cite{ren2022revisiting}.
To overcome this limitation, Fast-VGAN adopts an adversarial training approach. A discriminator is introduced during training to distinguish between real and synthesized Mel spectrograms. This encourages the generator to produce outputs that are perceptually closer to natural speech. Unlike diffusion-based models commonly used in recent work, we opt for a GAN-based approach to maintain faster inference times and reduce computational overhead during synthesis.
Fast-VGAN’s generator and discriminator are trained jointly in this adversarial setup. During inference, voice conversion can be achieved simply by replacing the speaker embedding with that of the target speaker. An illustration of the inference setup is provided in Figure~\ref{fig:fastvg-inf}.

\subsection{Input Features}

The input features were selected to enable voice conversion while preserving control over key speech characteristics. The model leverages four complementary modalities.

\vspace{0.25cm}

\noindent \textbf{Fundamental frequency} (F0), extracted using F0-FCN \cite{ardaillon2019fully}, captures prosodic information related to the speaker's pitch. The representation is expressed in the logarithmic domain, which aligns more closely with human pitch perception and helps normalize pitch variations across speakers. To further reduce the risk of encoding timbre-related information in the F0 signal, information that should ideally be captured by the speaker embedding, we subtract the speaker’s global log-mean F0. This normalization step encourages the model to rely on the dedicated speaker embedding for speaker-specific characteristics, while preserving the prosodic variation. \remove[AR]{The output is preprocessed by a simple Conv2D layer.} Please note that we do not apply the same normalization to the F0 standard deviation or ambitus: we assume that this F0 dynamics mostly encodes speaking style and expressivity of the speaker, while no timber-related information.

\vspace{0.25cm}

\noindent \textbf{Intensity}, computed as the average energy along the frequency axis of the Mel spectrogram, reflects the perceived loudness and expressivity of the utterance.

\vspace{0.25cm}
\noindent \textbf{Aligned phonemes}, obtained using the temporal alignment method proposed in \cite{teytaut2023temporal}, provide a structured representation of the linguistic content and articulation. Given a text file, we convert each sentence into a sequence of phonemes using automatic phonemization. The backend phonemizer \cite{black1997festival,bechet2001lia} is chosen according to the language. We represent these temporally aligned phonemes using an augmented one‑hot vector: alongside the usual one‑hot encoding, we append a value indicating the phoneme’s frame length. Then, the one-hot encoded labels are passed through a single Conv1D layer with kernel size 3, producing a learned embedding of the phoneme class, phoneme length, and phoneme context. Subsequently, these label embeddings are replicated according to the number of frames covered by each label. Finally, we concatenate a 2-dimensional positional encoding, designed to temporally localize each frame within the utterance. Each pair of additional dimensions will receive a sequence of continuous features consisting of linear cross-fades between 0 and 1. The first pair of dimensions receives a cross-fade that spans from the beginning to the end of the phrase and encodes the position of the frame within the phrase, and the second pair receives a sequence of cross-fades from the start to the end of each phoneme, encoding the position of the frame within the phoneme. These 4 dimensions ensure that the model has local access to \remove[AR]{the most} relevant information about the frame position.

\vspace{0.25cm}
\noindent \textbf{Speaker identity} Fast-VGAN has been configured in an any-to-many voice conversion setting. Accordingly, speaker identity is represented as a simple lookup embedding vector: each speaker in our dataset is assigned a unique, learnable vector; this table of vectors is initialized randomly and is jointly trained with the rest of the model\remove[AR]{, with no pretraining or external speaker encoder}. In the initial tests, we studied embeddings with and without a unit-norm constraint and found the unit-norm constraint beneficial. We kept it for the following experiments.

\vspace{0.25cm}
\change{Together, these features form a minimal and interpretable set of inputs enabling fine-grained control of speech prosody together with multi-speaker voice conversion in a compact and disentangled way.}{Together, these features form a minimal, disentangled, and interpretable set of conditioning parameters enabling fine-grained control of speech prosody and speaker identity. 
To create the input tensor, we replicate the speaker embedding over time, concatenate all features along the feature axis, replicate the result 5 times along the frequency axis, and apply a Conv2D layer with a (3,3) kernel.}

\subsection{Implementation details}

Table~\ref{tab:fastvg_decoder_discriminator} provides a detailed comparison of the convolutional configurations employed in both modules. 

\begin{table}[h]
\vspace{-0.3cm}
\centering
\resizebox{\columnwidth}{!}{
\begin{tabular}{lccc}
\toprule
 & \textbf{Decoder} & \textbf{2D Discriminator} & \textbf{1D Discriminator} \\
\midrule
Conv blocks & 5 (1 Conv2D-Tr   & 5 Conv2D & 4 Conv1D \\
 & + 3 Conv2D + ResBlock) & & \\
\vspace{-0.2cm} \\
Channels & [160, 144, 128, 112, 96] & [80, 100, 200, 300, 100] & [128, 128, 512, 128]\\
\vspace{0.2cm}
Kernel size & Conv2D:(3,3) & (3,3) & [(3,), (3,), (3,), (1,)] \\
\vspace{0.2cm}
Strides & [[1,2],[1,2],[1,2],[1,2],1] & [[2,2], [1,2], & 1 \\
 & & [1,1], [1,1], [1,1]] & \\
\vspace{0.2cm}
Activation & Swish & Leaky ReLU& Leaky ReLU \\
\bottomrule
\end{tabular}
}
\vspace{0.1cm}
\caption{Decoder and discriminator configuration. For stride and kernel specifications, the first dimension corresponds to the time axis and the second dimension to the frequency axis.}
\label{tab:fastvg_decoder_discriminator}
\vspace{-0.8cm}
\end{table}

\bigskip

The decoder is built with five convolutional blocks, each composed of a transposed Conv2D  layer, followed by three Conv2D layers and one residual block. The number of channels starts with 160 and is decreased in each transposed convolution (see Table~\ref{tab:fastvg_decoder_discriminator}). The final layer, not shown in Table~\ref{tab:fastvg_decoder_discriminator}, is a point-wise Conv2D layer with kernel size 1. The transposed convolutions have a kernel size that is multiplied by the stride. The decoder uses Swish activations and does not employ dropout or normalization layers, which, according to the experimental tests, do not improve the results.

The discriminator module is composed of two independent branches: a 2D convolutional discriminator and a 1D convolutional discriminator, which operate in parallel to assess different aspects of the generated spectrograms. The 2D branch focuses on local time-frequency patterns, while the 1D branch captures the global spectral structure over time. Both discriminators are conditioned on the target parameters, including F0 contour, speaker identity embedding, and phoneme labels. It uses Leaky ReLU activations and processes the input with gradually increasing channel sizes.

\section{Experiments}
\label{sec:experiments}

 In the remainder of this section, we present the databases and evaluation tasks performed. We first compare our method to two representative \textbf{voice conversion} models from the literature, \textit{ControlVC} \cite{chen2023controlvc} and \textit{HiFi-VC} \cite{kashkin2023hifi}, to benchmark the timbre transfer quality in a standard many-to-many conversion setting. We then analyze the role and variability of the input features used by our model, fundamental frequency, and phoneme alignment through a series of controlled experiments. The remainder of this section presents the experimental details of the evaluations.

\subsection{Datasets}

We conduct our experiments on two publicly available multi-speaker speech corpora. First, we use the \textbf{VCTK Corpus} \cite{yamagishi2019cstr}, containing approximately 44 hours of clean speech recorded from 110 English speakers, each uttering about 400 sentences. This one is used for the \textit{VC} and \textit{scaling} experiments. Second, we include the \textbf{Expresso Dataset} \cite{nguyen2023expresso}, composed of recordings from 4 English speakers (2 male, 2 female) expressing four emotions: neutral, confused, happy, and sad. This one is added to the VCTK Corpus for the \textit{expressive synthesis} experiments. While Fast-VGAN generates voice conversion at 24 kHz, we down-sampled all conversions to 16 kHz to ensure a fair comparison with the baselines during evaluation. All recordings from both datasets are resampled to 16 kHz.



\subsection{Training}


The Fast-VGAN generator is trained on the VCTK dataset using a single RTX 3080 GPU with a batch size of 32 for a total of 100k steps and 400 epochs ($\approx 5$ hours of training). Optimization is performed using the Adam optimizer with  learning rate $1e^{-4}$. We use mixed precision during training. When running inference on a CPU, the decoder (without vocoder) achieves a synthesis rate of more than 10 times faster than real-time, for generating individual phrases (batch-size=1), which is negligible compared to the vocoder, which is 2 times faster than real-time when generating 24 kHz.

\subsection{Experimental Setup}


We detail the configurations and evaluation protocols used across our four experiments: voice identity conversion benchmarking, voice identity conversion with adaptation of pitch and durations, static pitch shift and time-stretching and expressive synthesis without expressive training data.

\vspace{0.25cm}

\noindent \textbf{Voice identity conversion}: For voice identity conversion, we adopted an \change{any-to-any}{many-to-many} conversion setup. To do so, we selected a set of speakers who were present during training for all three systems. We chose 8 speakers from the VCTK dataset (4 female and 4 male) to perform the voice conversion evaluation. Among these 8 speakers, we selected 2 utterances that were spoken by all of them that were used for testing and removed from training. These samples are also retained for the subsequent experiments. 


\vspace{0.25cm}

\noindent \textbf{Voice identity conversion with adaptation of pitch and durations}:
While adapting the mean F0  to that of the target speaker is commonly used in VC \cite{bous2022voice}, we hypothesize here that the adaptation of the mean speech rate, but also the pitch and duration ambitus, play a role in the perception of a speaker's identity. Fast-VGAN makes it possible to control these parameters. We therefore propose a second identity conversion experiment, in which these parameters are adapted during conversion.
To do so, we extract the target speaker’s standard deviation of the mean F0 and apply it to modulate the source speaker’s f0 contour. We also estimate the target speaker’s average speech rate and apply a proportional time-stretching factor to the vowels in aligned phonemes of the source utterance. We evaluate each of these adaptations independently, as well as in combination.

\vspace{0.25cm}

\noindent \textbf{Static pitch shift and time-stretching}: 
Beyond voice conversion, Fast-VGAN offers the possibility to precisely control prosodic parameters, such as pitch and duration. We therefore evaluate to what extent those transformations remain natural and intelligible with respect to the transformation factor of the speech parameters. 
We apply pitch shifts of up to ±1 octave to evaluate the model's ability to transpose the voice while preserving naturalness. We also vary the vocal ambitus (F0 range) by up to ±1 octave to study the boundaries of expressive capacity. Finally, we stretch and compress the vowel durations by up to a factor of 3 to examine the model’s tolerance to extreme rate modifications. Note that we interpolate F0 and intensity curves to match the new aligned phonemes' length.

\vspace{0.25cm}

\noindent \textbf{Expressive Voice Conversion}: 
We propose an evaluation of the dynamic manipulation of these parameters, by applying pitch and duration variation curves. Unlike ControlVC, where dynamic modifications were chosen arbitrarily, we propose here to rely on a parallel emotion database like the Expresso dataset to evaluate the transfer of speech parameters to transform neutral speech into expressive speech.
To do so, we augmented the VCTK training data with four speakers from Expresso, using only their neutral utterances to allow the model to learn their timbre without exposure to expressive variations. At inference time, we use the expressive sentences from these same speakers to generate input features, allowing us to assess the model’s ability to synthesize expressive speech despite never being trained on expressive data. This setup investigates the generalization capabilities of the model for transferring prosodic and phonetic variation \cite{zhou2022speech, zhou2021seen}.

\section{Evaluations}
\label{sec:evaluations}



\subsection{Objective Evaluations}


For all conducted experiments, including the proposed method and the baseline systems, we evaluate the performance using both intelligibility and speaker similarity metrics. 
\begin{itemize}
    \item \textbf{Intelligibility}: we compute the Word Error Rate (WER) by transcribing all generated utterances using the pretrained \texttt{stt\_en\_fastconformer\_hybrid\_large\_pc} model available in the NeMo toolkit\footnote{\url{https://github.com/NVIDIA/NeMo}}. This model is based on a hybrid Conformer architecture trained on large-scale English datasets and offers state-of-the-art performance in ASR. 
    \item \textbf{Speaker similarity}: we employ the Resemblyzer library\footnote{\url{https://github.com/resemble-ai/Resemblyzer}}, which provides speaker embeddings computed from raw audio waveforms. For each generated sample, we extract its embedding and compare it to the embedding of the corresponding ground truth reference using cosine similarity. This metric quantifies how close the generated voice is to the intended target speaker in terms of timbre and vocal identity.
\end{itemize}
 All reported results include 95\% confidence intervals, computed via bootstrapping over the evaluation set.

\begin{figure*}[ht]
\centering

\begin{subfigure}[b]{0.32\textwidth}
    \includegraphics[width=\linewidth]{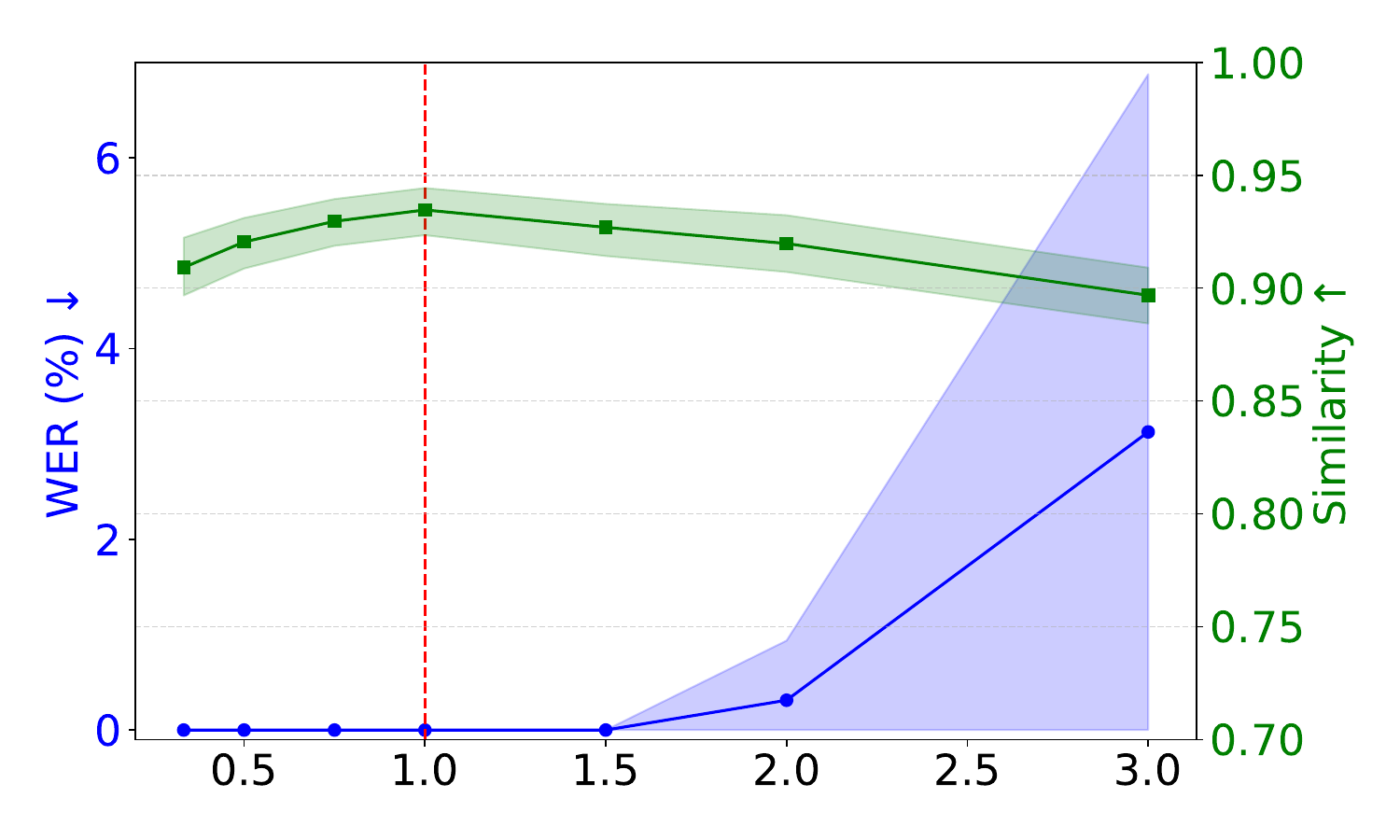}
    \caption{Vowel duration scaling}
    \label{fig:dilation}
\end{subfigure}
\hfill
\begin{subfigure}[b]{0.32\textwidth}
    \includegraphics[width=\linewidth]{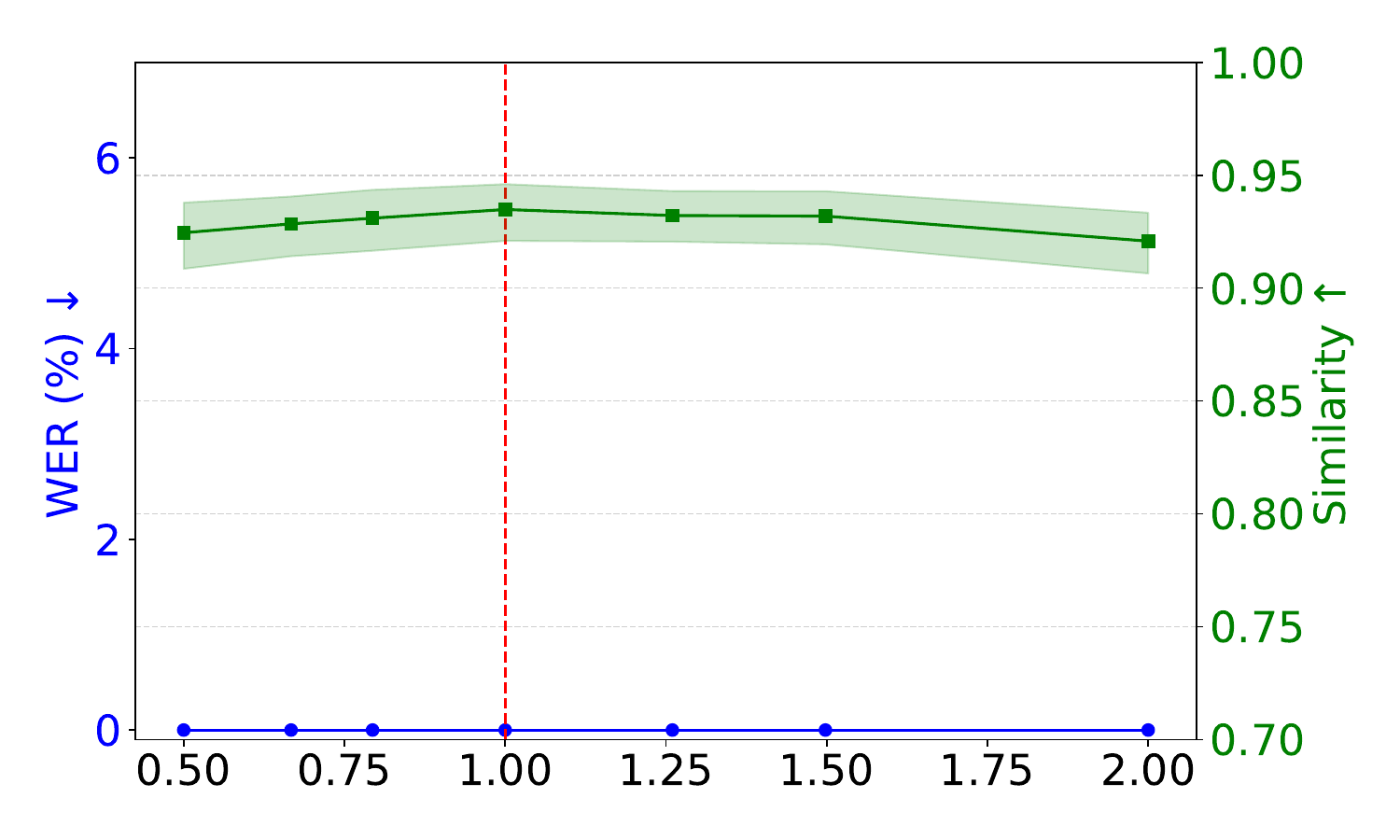}
    \caption{Ambitus scaling}
    \label{fig:ambitus}
\end{subfigure}
\hfill
\begin{subfigure}[b]{0.32\textwidth}
    \includegraphics[width=\linewidth]{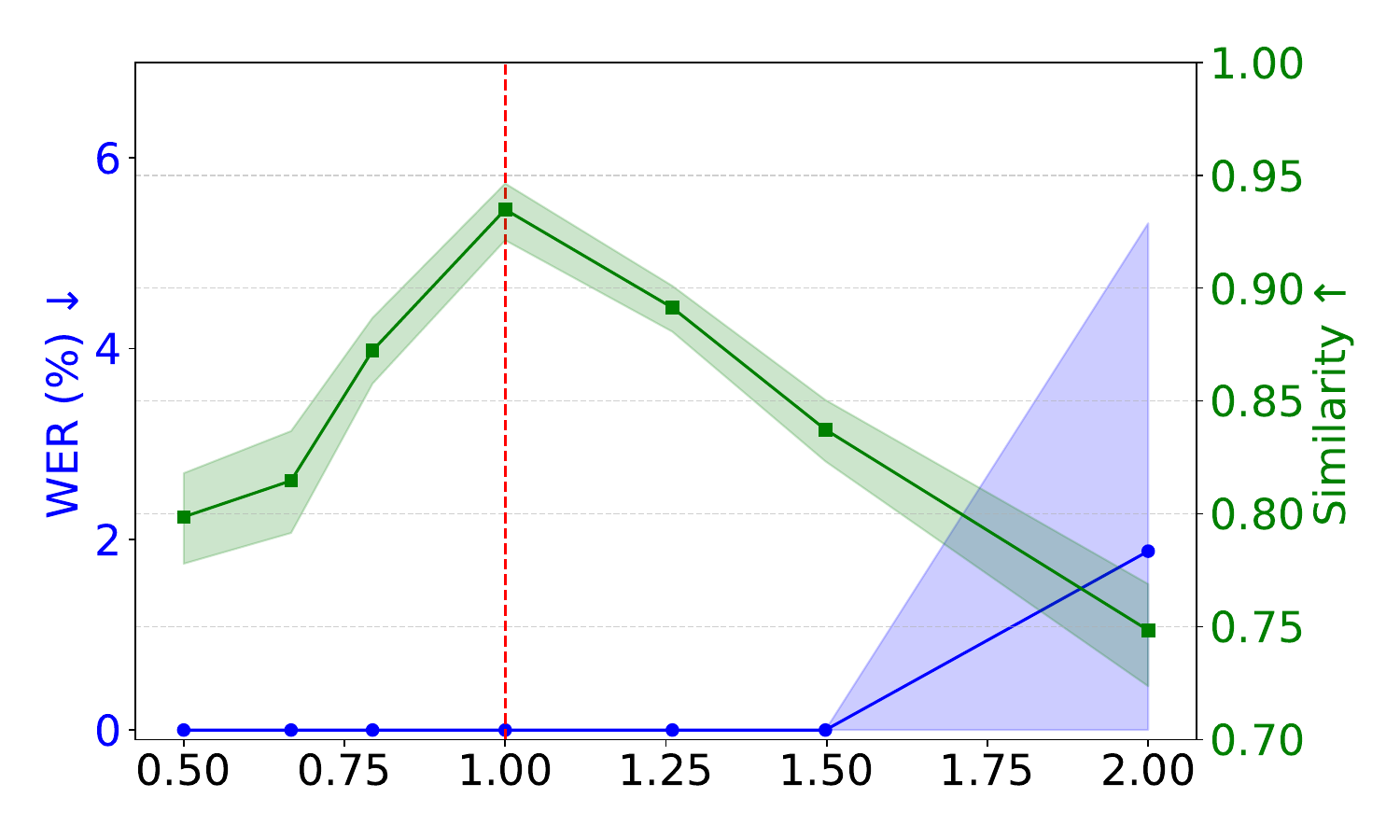}
    \caption{F0 shifting}
    \label{fig:f0}
\end{subfigure}

\caption{WER and speakersimilarity under prosodic parameter scaling. Vertical dashed line indicates the neutral setting. WER is shown as a percentage (0–100), similarity is normalized between 0 and 1.}
\label{fig:scaling_experiments}
\vspace{-0.4cm}
\end{figure*}

\subsubsection{Voice identity conversion with parameters adaptation}

Table \ref{tab:vc_results} shows the results of voice conversion on seen speakers during training.

\begin{table}[h]
\vspace{-0.2cm}
\centering
\resizebox{\columnwidth}{!}{
\begin{tabular}{lcc}
\toprule
Model \& num. params & \textbf{WER (\% - min: 0.) $\downarrow$} & \textbf{Sim. (max: 1.) $\uparrow$} \\
\midrule
ControlVC ($\approx 20$M) & $0.089 \pm 0.112$ & $\bm{0.652 \pm 0.014}$ \\

HifiVC ($\approx 14$M) & $2.857 \pm 0.738$ & $0.586 \pm 0.009$ \\

Fast-VGAN ($\approx 3.2$M) & $\bm{0.000 \pm 0.000}$ & $0.648 \pm 0.016$ \\

Fast-VGAN Amb & $0.446 \pm 0.425$ & $0.649 \pm 0.016$ \\

Fast-VGAN Dil & $\bm{0.000 \pm 0.000}$ & $0.645 \pm 0.016$ \\

Fast-VGAN Amb \& Dil & $0.357 \pm 0.446$ & $0.646 \pm 0.015$ \\

\bottomrule
\end{tabular}
}
\caption{Objective VC evaluation. Amb : target speaker's ambitus applied, Dil : target speaker's speech rate applied.}
\label{tab:vc_results}
\vspace{-0.6cm}
\end{table}

Note that for ground truth, WER and similarity are $0.00\pm0.00\%$ and $1.00\pm0.00\%$. Among the baselines, ControlVC shows low WER and good similarity. Our proposed method achieves a WER of 0.00, indicating excellent intelligibility preservation, and a competitive similarity score. When adding target speaker-specific features, such as ambitus and speech rate dilation, results remain stable in both intelligibility and speaker similarity, with a slight improvement in similarity when adding ambitus. These results confirm that Fast-VGAN not only outperforms the baselines in WER but also maintains high speaker similarity, and that the conditioning on speaker-specific prosodic features does not degrade the model’s performance. We also note that WER for Fast-VGAN only doing resynthesis and the standalone MBExWN vocoder are $0.00\pm0.00\%$ and $0.935 \pm 0.013$ and $0.992 \pm 0.001$ respectively for similarity.

\subsubsection{Static pitch shift and time-stretching}

To assess the controllability and robustness of our system under extreme prosodic variations, we designed three separate experiments focusing on different aspects of voice modulation. The results are shown in Figure \ref{fig:scaling_experiments}. We explore: (i) vowel duration scaling (fig \ref{fig:dilation}), where durations are compressed up to 3x or stretched down to $\frac{1}{3}$x; vocal range manipulation (fig \ref{fig:ambitus}), where the ambitus is either increased or decreased by up to one octave; and (iii) F0 transposition (fig \ref{fig:f0}) where the mean pitch of the source speaker is shifted up or down by one octave. In all plots, we report both the WER and the speaker similarity, with confidence intervals, and mark the neutral reference point (i.e., no transformation applied) using a vertical red dashed line. These results show that our model preserves intelligibility and speaker consistency across a wide range of expressive transformations, making it suitable for high-level control of speech generation without compromising output quality.

\subsubsection{Expressive synthesis}

We conducted the same objective evaluations as in the VC benchmarking section, this time on the Expresso dataset. Table~\ref{tab:expr_results} reports the scores between input and output transcriptions. Metrics are presented globally as well as per emotion, for both the original (real) and synthesized (Fast-VGAN) utterances. Although WER scores remain relatively high, even for the real recordings, our expressive resynthesis model does not degrade intelligibility. Fast-VGAN achieves comparable WERs to original speech, indicating strong preservation of linguistic content. Similarity scores for Fast-VGAN remain consistently high across all emotions, demonstrating the model’s ability to accurately keep the speaker identity, even in a setting where such expressive styles were unseen during training.

\begin{table}[h]
\vspace{0cm}
\centering
\resizebox{\columnwidth}{!}{
\begin{tabular}{lcc}
\toprule
 & \textbf{WER (\% - min: 0.) $\downarrow$} & \textbf{Sim. (max: 1.) $\uparrow$} \\
\midrule
Real global & $19.38 \pm 7.13$ & $1.000 \pm 0.000$ \\
Confused & $12.80 \pm 6.67$ & - \\
Happy & $18.07 \pm 8.90$ & - \\
Sad & $16.18 \pm 6.46$ &  - \\
\midrule
Fast-VGAN global & $19.46 \pm 4.75$ & $0.867 \pm 0.029$ \\
Confused & $18.57 \pm 8.88$ & $0.825 \pm 0.045$ \\
Happy & $16.48 \pm 5.55$ & $0.895 \pm 0.022$ \\
Sad & $16.07 \pm 9.68$ & $0.884 \pm 0.034$ \\
\bottomrule
\end{tabular}
}
\vspace{0.1cm}
\caption{Objective expressive synthesis evaluation.}
\label{tab:expr_results}
\vspace{-0.9cm}
\end{table}

\begin{figure*}[ht]
\centering
\begin{subfigure}[b]{\columnwidth}
    \includegraphics[width=0.9\linewidth]{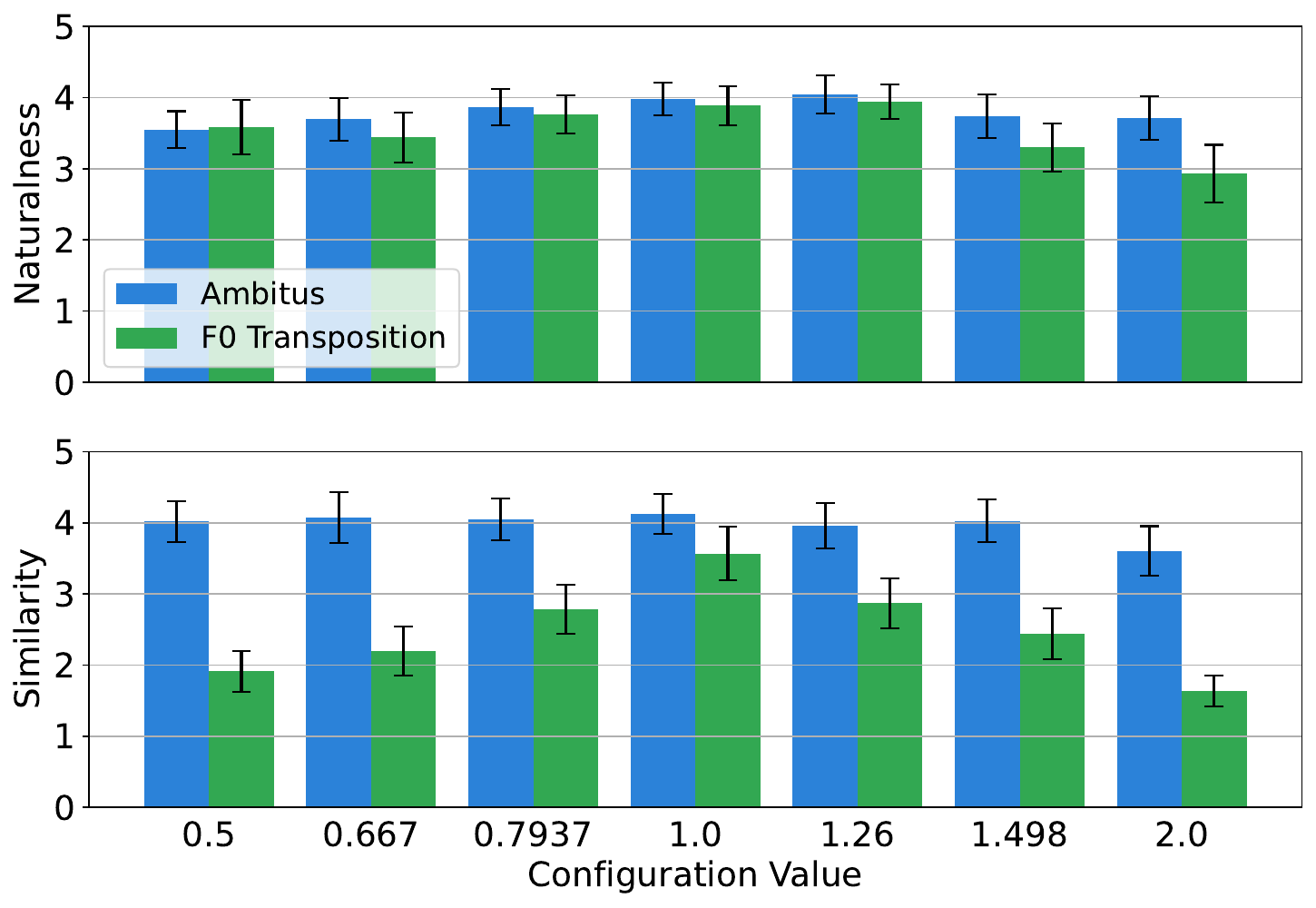}
    \caption{Ambitus scaling and F0 shifting}
    \label{fig:mos_f0}
\end{subfigure}
\hfill
\begin{subfigure}[b]{\columnwidth}
    \includegraphics[width=0.9\linewidth]{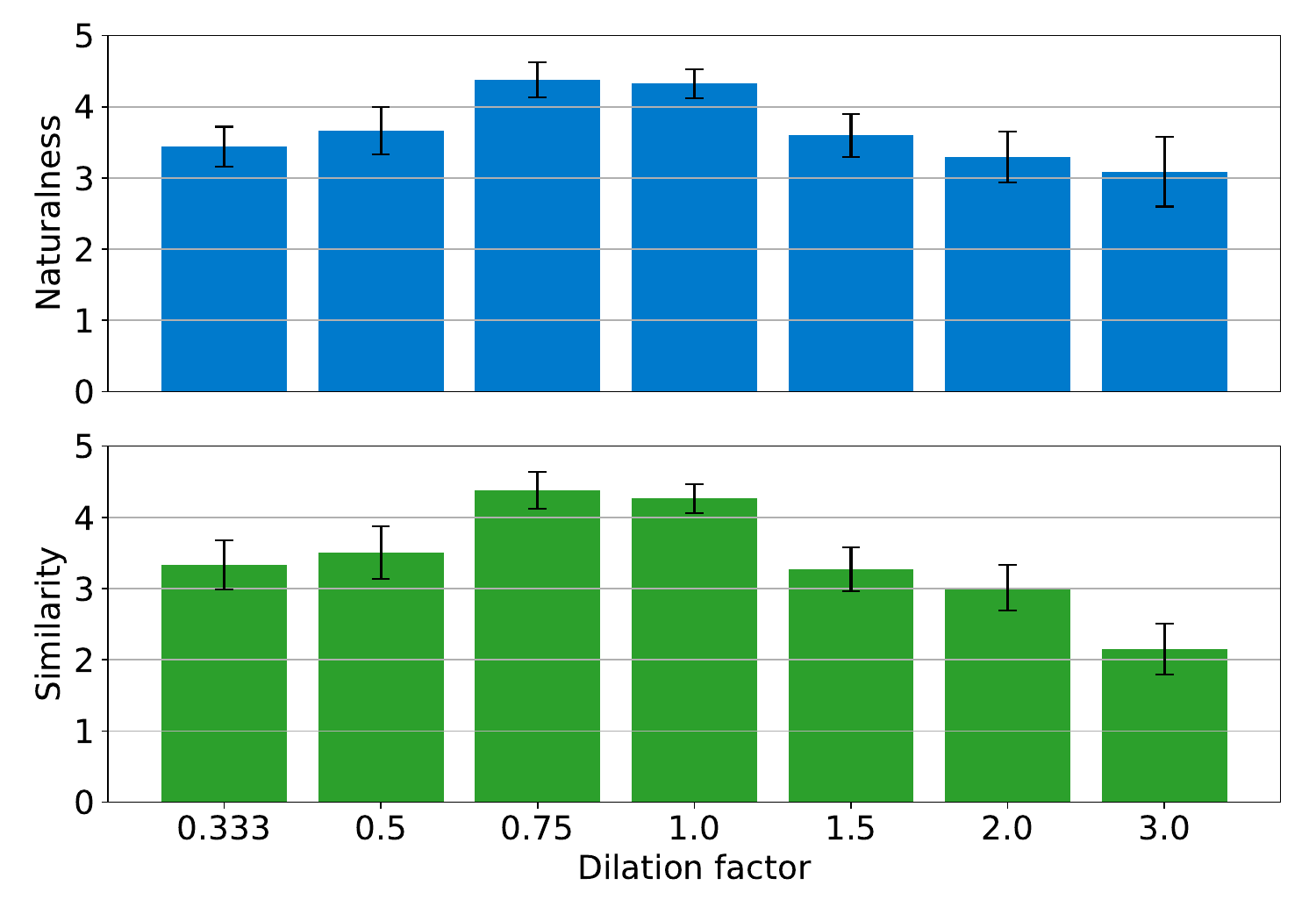}
    \caption{Vowel duration scaling}
    \label{fig:mos_dil}
\end{subfigure}
\vspace{-0.2cm}
\caption{Subjective parameter scaling evaluation.}
\label{fig:mos_res}
\vspace{-0.4cm}
\end{figure*}

\subsection{Subjective Evaluation}

To assess the perceptual quality of the proposed system, we conducted a series of Mean Opinion Score (MOS) evaluations using the Prolific\footnote{\url{https://www.prolific.com/}} crowdsourcing platform. Participants were asked to rate audio samples along two axes: naturalness — how natural or synthetic the voice sounds — and speaker similarity — how close the converted voice sounds to the target speaker. Each attribute was rated on a 5-point Likert scale, ranging from 1 (bad) to 5 (excellent). The main voice conversion experiment, comparing Fast-VGAN, Hifi-VC and ControlVC to real target speech, involved 40 unique participants. Additional experiments focused on prosodic control — including \emph{F0 transposition}, \emph{ambitus scaling}, and \emph{vowel dilation} — and experiment on expressive synthesis were evaluated independently, each by approximately 20 participants.


  

\subsubsection{Voice identity conversion with parameter adaptation}

Table~\ref{tab:mos_results} presents the results of the subjective MOS evaluation for naturalness and speaker similarity. 
\begin{table}[h]
\centering
\resizebox{\columnwidth}{!}{
\begin{tabular}{lcc}
\toprule
 & \textbf{Naturalness (0-5) $\uparrow$} & \textbf{Similarity (0-5) $\uparrow$} \\
\midrule
Ground truth & $4.35 \pm 0.12$ & $4.34 \pm 0.17$ \\

ControlVC & $3.60 \pm 0.14$ & $2.82 \pm 0.18$ \\

HifiVC & $2.96 \pm 0.19$ & $2.00 \pm 0.17$ \\

Fast-VGAN & $\bm{3.63 \pm 0.25}$ & $\bm{3.47 \pm 0.32}$ \\

Fast-VGAN Amb & $3.53 \pm 0.35$ & $3.15 \pm 0.42$ \\

Fast-VGAN Dil & $3.62 \pm 0.28$ & $3.34 \pm 0.35$ \\

Fast-VGAN Amb \& Dil & $3.13 \pm 0.31$ & $2.99 \pm 0.34$ \\

\bottomrule
\end{tabular}
}
\vspace{0.1cm}
\caption{Subjective VC evaluation. Amb: target speaker's ambitus applied, Dil: target speaker's speech rate applied.}
\label{tab:mos_results}
\vspace{-4mm}
\end{table}
%
%
Overall, our proposed system achieves the highest performance across both dimensions, with a naturalness score of $3.63 \pm 0.25$ and a speaker similarity score of $3.47 \pm 0.32$. This suggests that the model is capable of producing high-quality, expressive speech while effectively preserving the identity of the target speaker. ControlVC shows strong performance in terms of naturalness ($3.60 \pm 0.14$), but is outperformed by Fast-VGAN in speaker similarity. In contrast, HiFi-VC underperforms in both metrics, particularly in similarity ($2.00 \pm 0.17$). Applying only the target speaker's ambitus or only vowel dilation results in slight decreases in similarity but maintains comparable naturalness. Notably, Fast-VGAN Dil achieves a similarity score of $3.34 \pm 0.35$, very close to the original Fast-VGAN model, indicating that modifying speech rate has a limited adverse effect on perceived identity. While we hypothesized that adapting prosodic parameters during conversion would enhance the perception of speaker identity, the results suggest otherwise. When both ambitus and dilation are applied simultaneously, we observe a drop in both naturalness and similarity. This suggests that aggressive or combined prosodic manipulations may introduce artifacts or degrade speaker identity, emphasizing the need for careful parameter calibration in expressive applications.

\vspace{-0.2cm}
\subsubsection{Static pitch shift and time-stretching}

We progressively modified one prosodic dimension at a time: vowel duration (dilation), pitch range (ambitus), and F0 transposition. As illustrated in Figure~\ref{fig:mos_res}, we observe bell-shaped curves across all configurations, with MOS scores (both for naturalness and speaker similarity) peaking when the scaling factor is close to the neutral value (i.e., a factor of 1). This pattern confirms that extreme modifications tend to degrade perceived quality and identity, yet also highlights the resilience of our model to moderate prosodic deviations. Notably, the ambitus and dilation transformations maintain relatively high MOS scores even at more extreme values (e.g., up to $0.5\times$ or $2.0\times$), suggesting that these parameters can be safely used to introduce expressive variation without strongly affecting the perceived speaker identity. This is consistent with the fact that pitch range and speech rate are generally more flexible and less tightly coupled to speaker identity than absolute F0. In contrast, F0 transposition—being directly linked to perceived pitch height—leads to more pronounced drops in similarity scores at higher scaling factors, as expected.

\vspace{-0.2cm}
\subsubsection{Expressive synthesis}

\begin{figure}[h]
\vspace{-0.3cm}
    \centering
    \includegraphics[width=1\linewidth]{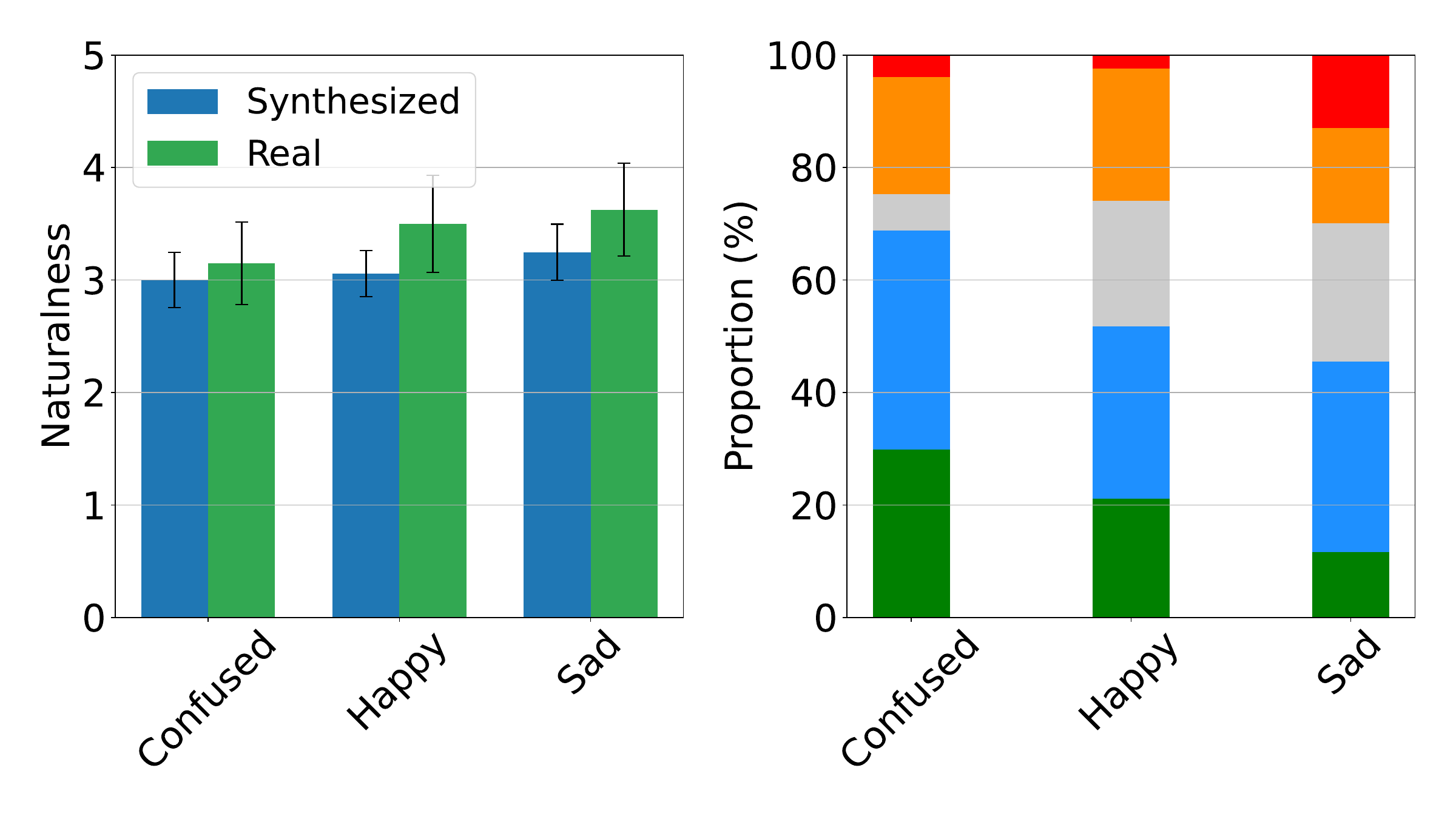}
    \vspace{-0.9cm}
    \caption{Naturalness over resynthesized and real Expr. audio and similarity to targeted emotion. Green: target, blue: target (fairly), grey: uncertain, orange: neutral (fairly), red: neutral.}
    \vspace{-0.3cm}
    \label{fig:expr_mos}
\vspace{-0.5cm}
\end{figure}


In this experiment, participants had to judge real and converted speech utterances expressing various target emotions (\emph{confused}, \emph{happy}, \emph{sad}). They were asked to rate their naturalness, and to judge whether the converted emotion is more similar to neutral reference or to a reference of the target emotion, on a 5-degree scale. Figure~\ref{fig:expr_mos} shows the results, comparing the naturalness of real and converted speech and the proportion of trials where the converted emotion was judged more similar to the target emotion reference. We observe that applying dynamic contours does not substantially degrade perceived naturalness compared to real recordings, with scores in the range of those reported for static modification. Informal listening to the converted speech indicates that the dynamic modification of pitch and duration with Fast-VGAN can successfully modify the other timbre-related acoustic cues accordingly, which results into more realistic emotion conversion. About the similarity to the reference, results are encouraging but nuanced: conversion to \emph{confused} were most consistently rated as closer to the target emotion. In contrast, \emph{happy} and \emph{sad} yielded more mixed outcomes: while most conversions were seen as more similar to the target, a notable share of uncertain judgments emerged. This may reflect from the inherent ambiguity of vocal emotion perception, as reported in~\cite{robinson2019}.

\vspace{-0.3cm}

\section{Conclusion}
\label{sec:conclusion}
\vspace{-0.2cm}
Fast-VGAN demonstrates strong potential for expressive voice conversion, enabling controllable transformations of prosodic features such as pitch range and speech rate while maintaining naturalness and speaker similarity. Our results suggest that explicit prosodic conditioning can effectively drive perceived expressiveness and provide users with meaningful, interpretable control over synthesized speech. Moreover, the model achieves competitive performance without requiring expressive training data, highlighting the efficiency of our prosody-guided approach. Future work includes the integration of speaker encoding mechanisms to move beyond our current many-to-many setup and enable true any-to-any voice conversion, where both source and target speakers may be unseen during training.

\section{Acknowledgments}

This work was partly performed using HPC resources from GENCI-IDRIS (Grant 2025-AD011011177R5), and partly funded by the ANR project BRUEL (ANR-22-CE39-0009).

\bibliographystyle{IEEEtran}

\begin{thebibliography}{10}
\providecommand{\url}[1]{#1}
\csname url@samestyle\endcsname
\providecommand{\newblock}{\relax}
\providecommand{\bibinfo}[2]{#2}
\providecommand{\BIBentrySTDinterwordspacing}{\spaceskip=0pt\relax}
\providecommand{\BIBentryALTinterwordstretchfactor}{4}
\providecommand{\BIBentryALTinterwordspacing}{\spaceskip=\fontdimen2\font plus
\BIBentryALTinterwordstretchfactor\fontdimen3\font minus \fontdimen4\font\relax}
\providecommand{\BIBforeignlanguage}[2]{{%
\expandafter\ifx\csname l@#1\endcsname\relax
\typeout{** WARNING: IEEEtran.bst: No hyphenation pattern has been}%
\typeout{** loaded for the language `#1'. Using the pattern for}%
\typeout{** the default language instead.}%
\else
\language=\csname l@#1\endcsname
\fi
#2}}
\providecommand{\BIBdecl}{\relax}
\BIBdecl

\bibitem{bargum2024reimagining}
A.~R. Bargum, S.~Serafin, and C.~Erkut, ``{Reimagining speech: a scoping review of deep learning-based methods for non-parallel voice conversion},'' \emph{Frontiers in Signal Processing}, vol.~4, 2024.

\bibitem{sisman2020overview}
B.~Sisman, J.~Yamagishi, S.~King, and H.~Li, ``{An Overview of Voice Conversion and Its Challenges: From Statistical Modeling to Deep Learning},'' \emph{IEEE/ACM Transactions on Audio, Speech, and Language Processing}, vol.~29, pp. 132--157, 2020.

\bibitem{qian2019autovc}
K.~Qian, Y.~Zhang, S.~C. Chang \emph{et~al.}, ``{AutoVC: Zero-Shot Voice Style Transfer with Only Autoencoder Loss},'' in \emph{International Conference on Machine Learning (ICML)}, 2019, pp. 5210--5219.

\bibitem{popov2022diffvc}
V.~Popov, I.~Vovk, V.~Gogoryan, T.~Sadekova, M.~S. Kudinov, and J.~Wei, ``{Diffusion-Based Voice Conversion with Fast Maximum Likelihood Sampling Scheme},'' in \emph{International Conference on Learning Representation (ICLR)}, 2022.

\bibitem{guo2023quickvc}
H.~Guo, C.~Liu \emph{et~al.}, ``{QUICKVC: A Lightweight VITS-Based Any-to-Many Voice Conversion Model using ISTFT for Faster Conversion},'' in \emph{IEEE Automatic Speech Recognition and Understanding Workshop (ASRU)}, 2023, pp. 1--7.

\bibitem{bous2022voice}
F.~Bous, L.~Benaroya, N.~Obin, and A.~Roebel, ``{Voice Reenactment with F0 and timing constraints and adversarial learning of conversions},'' in \emph{European Signal Processing Conference (EUSIPCO)}, 2022, pp. 389--393.

\bibitem{byun2023highlycontrollablediffvc}
\BIBentryALTinterwordspacing
K.~Byun, S.~Moon, and E.~Visser, ``{Highly Controllable Diffusion-based Any-to-Any Voice Conversion Model with Frame-level Prosody Feature},'' 2023. [Online]. Available: \url{https://arxiv.org/abs/2309.03364}
\BIBentrySTDinterwordspacing

\bibitem{kashkin2023hifi}
A.~Kashkin, I.~Karpukhin, and S.~Shishkin, ``{HiFi-VC: High Quality ASR-based Voice Conversion},'' in \emph{Speech Synthesis Workshop (SSW)}, 2023, pp. 100--105.

\bibitem{chen2023controlvc}
M.~Chen and Z.~Duan, ``{ControlVC: Zero-Shot Voice Conversion with Time-Varying Controls on Pitch and Speed},'' in \emph{Interspeech}, 2023, pp. 2098--2102.

\bibitem{campbell2003voice}
N.~Campbell and P.~Mokhtari, ``{Voice quality: the 4th prosodic dimension},'' in \emph{International Congress of Phonetic Sciences (ICPhS)}, 2003, pp. 2417--2420.

\bibitem{benaroyavc2923}
L.~Benaroya, N.~Obin, and A.~Roebel, ``{Manipulating Voice Attributes by Adversarial Learning of Structured Disentangled Representations},'' \emph{Entropy}, vol.~25, no.~2, 2023.

\bibitem{donahue2021endtoend}
J.~Donahue, S.~Dieleman, M.~Binkowski, E.~Elsen, and K.~Simonyan, ``{End-to-End Adversarial Text-to-Speech},'' in \emph{International Conference on Learning Representation (ICLR)}, 2021.

\bibitem{gengembre2024disentangling}
N.~Gengembre, O.~Le~Blouch, and C.~Gendrot, ``{Disentangling prosody and timbre embeddings via voice conversion},'' in \emph{Interspeech}, 2024, pp. 2765--2769.

\bibitem{deng2024learning}
Y.~Deng, J.~Wang, X.~Zhang, N.~Cheng, and J.~Xiao, ``{Learning Expressive Disentangled Speech Representations with Soft Speech Units and Adversarial Style Augmentation},'' in \emph{International Joint Conference on Neural Networks (IJCNN)}.\hskip 1em plus 0.5em minus 0.4em\relax IEEE, 2024, pp. 1--7.

\bibitem{qu2023disentangling}
L.~Qu, T.~Li, C.~Weber, T.~Pekarek-Rosin, F.~Ren, and S.~Wermter, ``{Disentangling prosody representations with unsupervised speech reconstruction},'' \emph{IEEE/ACM Transactions on Audio, Speech, and Language Processing}, vol.~32, pp. 39--54, 2023.

\bibitem{sivaprasad2021emotional}
S.~Sivaprasad, S.~Kosgi, and V.~Gandhi, ``{Emotional prosody control for speech generation},'' \emph{arXiv preprint arXiv:2111.04730}, 2021.

\bibitem{rana2024advancements}
P.~S. Rana, S.~K. Modi, A.~L. Yadav \emph{et~al.}, ``Advancements in real-time voice conversion technologies: A comprehensive analysis of techniques,'' \emph{Chahat and Modi, Soham Kr and Yadav, Anup Lal, Advancements in Real-Time Voice Conversion Technologies: A Comprehensive Analysis of Techniques (August 20, 2024)}, 2024.

\bibitem{lee2019robust}
Y.~Lee and T.~Kim, ``{Robust and fine-grained prosody control of end-to-end speech synthesis},'' in \emph{ICASSP 2019-2019 IEEE International Conference on Acoustics, Speech and Signal Processing (ICASSP)}.\hskip 1em plus 0.5em minus 0.4em\relax IEEE, 2019, pp. 5911--5915.

\bibitem{tjandra2020unsupervised}
A.~Tjandra, R.~Pang, Y.~Zhang, and S.~Karita, ``{Unsupervised learning of disentangled speech content and style representation},'' \emph{arXiv preprint arXiv:2010.12973}, 2020.

\bibitem{wang2020learning}
D.~Wang, S.~Liu, L.~Sun, X.~Wu, X.~Liu, and H.~Meng, ``Learning explicit prosody models and deep speaker embeddings for atypical voice conversion,'' \emph{arXiv preprint arXiv:2011.01678}, 2020.

\bibitem{gulati2020conformer}
A.~Gulati, J.~Qin, C.-C. Chiu, N.~Parmar, Y.~Zhang, J.~Yu, W.~Han, S.~Wang, Z.~Zhang, Y.~Wu, and R.~Pang, ``{Conformer: Convolution-augmented Transformer for Speech Recognition},'' in \emph{Interspeech}, 2020, pp. 5036--5040.

\bibitem{morise2016world}
M.~Morise, F.~Yokomori, and K.~Ozawa, ``{WORLD: A Vocoder-Based High-Quality Speech Synthesis System for Real-Time Applications},'' in \emph{IEICE Transactions on Information and Systems}, vol. E99.D, no.~7, 2016, pp. 1877--1884.

\bibitem{huang2021far}
T.-h. Huang, J.-h. Lin, and H.-y. Lee, ``{How far are we from robust voice conversion: A survey},'' in \emph{2021 IEEE spoken language technology workshop (SLT)}.\hskip 1em plus 0.5em minus 0.4em\relax IEEE, 2021, pp. 514--521.

\bibitem{kong2020hifi}
J.~Kong, J.~Kim, and J.~Bae, ``{HiFi-GAN: Generative Adversarial Networks for Efficient and High Fidelity Speech Synthesis},'' in \emph{Proc. NeurIPS}, 2020, pp. 17\,022--17\,033.

\bibitem{du2021disentanglement}
Z.~Du, B.~Sisman, K.~Zhou, and H.~Li, ``Disentanglement of emotional style and speaker identity for expressive voice conversion,'' \emph{arXiv preprint arXiv:2110.10326}, 2021.

\bibitem{charpentier1986diphone}
F.~Charpentier and M.~Stella, ``{Diphone synthesis using an overlap-add technique for speech waveforms concatenation},'' in \emph{International Conference on Acoustics, Speech, and Signal Processing (ICASSP)}, 1986, pp. 2015--2018.

\bibitem{hsu2021hubert}
W.-N. Hsu, B.~Bolte, Y.-H.~H. Tsai, K.~Lakhotia, R.~Salakhutdinov, and A.~Mohamed, ``{HuBERT: Self-Supervised Speech Representation Learning by Masked Prediction of Hidden Units},'' in \emph{IEEE/ACM Transactions on Audio, Speech, and Language Processing}, vol.~29, 2021, pp. 3451--3460.

\bibitem{baevski2020wav2vec}
A.~Baevski, Y.~Zhou, A.~Mohamed, and M.~Auli, ``{wav2vec 2.0: A framework for self-supervised learning of speech representations},'' in \emph{NeurIPS}, 2020.

\bibitem{morise2009fast}
M.~Morise, H.~Kawahara, and H.~Katayose, ``{Fast and reliable f0 estimation method based on the period extraction of vocal fold vibration of singing voice and speech},'' in \emph{Audio Engineering Society Conference: 35th International Conference: Audio for Games}.\hskip 1em plus 0.5em minus 0.4em\relax Audio Engineering Society, 2009.

\bibitem{kim2018crepe}
J.~W. Kim, J.~Salamon, P.~Li, and J.~P. Bello, ``{Crepe: A Convolutional Representation for Pitch Estimation},'' in \emph{EEE International Conference on Acoustics, Speech and Signal Processing (ICASSP)}, 2018, pp. 161--165.

\bibitem{ardaillon2019fully}
L.~Ardaillon and A.~Roebel, ``{Fully-Convolutional Network for Pitch Estimation of Speech Signals},'' in \emph{Insterspeech}, 2019.

\bibitem{kaneko2017parallel}
T.~Kaneko and H.~Kameoka, ``{Parallel-Data-Free Voice Conversion Using Cycle-Consistent Adversarial Networks},'' in \emph{IEEE transactions on knowledge and data engineering}, 2017.

\bibitem{ferrovc2020}
R.~Ferro, N.~Obin, and A.~Roebel, ``{CycleGAN Voice Conversion of Spectral Envelopes using Adversarial Weights},'' in \emph{European Signal Processing Conference (EUSIPCO)}, 2020.

\bibitem{kaneko2019starganvc}
T.~Kaneko and H.~Kameoka, ``{StarGAN-VC: Non-parallel many-to-many voice conversion using star generative adversarial networks},'' in \emph{Spoken Language Technology Workshop (SLT)}, 2019, pp. 266--273.

\bibitem{kaneko2020starganv2}
------, ``{StarGAN-VC2: Rethinking conditional methods for stargan-based voice conversion},'' in \emph{Interspeech}, 2019, pp. 679--683.

\bibitem{qian2020nvcnet}
B.~Nguyen and F.~Cardinaux, ``{Nvc-net: End-to-end adversarial voice conversion},'' in \emph{ICASSP 2022-2022 IEEE International Conference on Acoustics, Speech and Signal Processing (ICASSP)}.\hskip 1em plus 0.5em minus 0.4em\relax IEEE, 2022, pp. 7012--7016.

\bibitem{tang2022avqvc}
H.~Tang, X.~Zhang, J.~Wang, N.~Cheng, and J.~Xiao, ``{Avqvc: One-shot voice conversion by vector quantization with applying contrastive learning},'' in \emph{ICASSP 2022-2022 IEEE International Conference on Acoustics, Speech and Signal Processing (ICASSP)}.\hskip 1em plus 0.5em minus 0.4em\relax IEEE, 2022, pp. 4613--4617.

\bibitem{popov2021grad}
V.~Popov, A.~Vovk, I.~Yakovenko \emph{et~al.}, ``{Grad-TTS: A Diffusion Probabilistic Model for Text-to-Speech},'' in \emph{International Conference on Machine Learning (ICML)}, 2021.

\bibitem{choi2024dddm}
H.-Y. Choi, S.-H. Lee, and S.-W. Lee, ``{DDDM-VC: Decoupled Denoising Diffusion Models with Disentangled Representation and Prior Mixup for Verified Robust Voice Conversion},'' in \emph{Proceedings of the AAAI Conference on Artificial Intelligence}, vol.~38, no.~16, 2024, pp. 17\,862--17\,870.

\bibitem{roebel2022neural}
A.~Roebel and F.~Bous, ``{Neural Vocoding for Singing and Speaking Voices with the Multi-Band Excited WaveNet},'' \emph{Information}, vol.~13, no.~3, p. 103, 2022.

\bibitem{ren2022revisiting}
Y.~Ren, X.~Tan, T.~Qin, Z.~Zhao, and T.-Y. Liu, ``{Revisiting Over-Smoothness in Text to Speech},'' in \emph{Proceedings of the 60th Annual Meeting of the Association for Computational Linguistics (ACL)}, 2022, pp. 8197--8213.

\bibitem{teytaut2023temporal}
Y.~Teytaut, ``{On temporal constraints for deep neural voice alignment},'' Ph.D. dissertation, Sorbonne Universit{\'e}, 2023.

\bibitem{black1997festival}
A.~Black, ``{Festival speech synthesis system: system documentation (1.1. 1)},'' \emph{Human Communication Research Centre Technical Report}, 1997.

\bibitem{bechet2001lia}
F.~B{\'e}chet, ``{LIA―PHON: Un syst{\`e}me complet de phon{\'e}tisation de textes},'' \emph{TAL. Traitement automatique des langues}, vol.~42, no.~1, pp. 47--67, 2001.

\bibitem{yamagishi2019cstr}
J.~Yamagishi, C.~Veaux, K.~MacDonald \emph{et~al.}, ``{CSTR VCTK Corpus: English multi-speaker corpus for CSTR voice cloning toolkit (version 0.92)},'' \emph{University of Edinburgh. The Centre for Speech Technology Research (CSTR)}, pp. 271--350, 2019.

\bibitem{nguyen2023expresso}
T.~A. Nguyen, W.-N. Hsu, A.~D'Avirro, B.~Shi, I.~Gat, M.~Fazel-Zarani, T.~Remez, J.~Copet, G.~Synnaeve, M.~Hassid, F.~Kreuk, Y.~Adi, and E.~Dupoux, ``{EXPRESSO: A Benchmark and Analysis of Discrete Expressive Speech Resynthesis},'' in \emph{Interspeech}, 2023, pp. 4823--4827.

\bibitem{zhou2022speech}
K.~Zhou, B.~Sisman, R.~Rana, B.~W. Schuller, and H.~Li, ``{Speech Synthesis with Mixed Emotions},'' \emph{IEEE Transactions on Affective Computing}, vol.~14, no.~4, pp. 3120--3134, 2022.

\bibitem{zhou2021seen}
K.~Zhou, B.~Sisman, R.~Liu, and H.~Li, ``{Seen and unseen emotional style transfer for voice conversion with a new emotional speech dataset},'' in \emph{IEEE International Conference on Acoustics, Speech and Signal Processing (ICASSP)}, 2021, pp. 920--924.

\bibitem{robinson2019}
C.~Robinson, N.~Obin, and A.~Roebel, ``{Sequence-to-Sequence Moelling OF F0 For Speech Emotion Conversion},'' in \emph{{IEEE International Conference on Acoustics, Speech, and Signal Processing (ICASSP)}}, Brighton, United Kingdom, May 2019.

\end{thebibliography}

\end{document}